\documentclass[aip,
amsmath,amssymb,
reprint,
]{revtex4-1}
\usepackage{graphicx}
\usepackage{siunitx}
\usepackage{chemformula}
\usepackage{xcolor}
\begin{document}

\title{Fast Fourier transform and multi-Gaussian fitting of XRR data to determine the thickness of ALD grown thin films within the initial growth regime}

\author{Michaela Lammel}
\email[Corresponding author: ]{m.lammel@ifw-dresden.de}
\affiliation{Institute for Metallic Materials, Leibniz Institute of Solid State and Materials Science, 01069 Dresden, Germany}
\affiliation{Technische Universit\"at Dresden, Institute of Applied Physics, 01062 Dresden, Germany}

\author{Kevin Geishendorf}%
\affiliation{Institute for Metallic Materials, Leibniz Institute of Solid State and Materials Science, 01069 Dresden, Germany}
\affiliation{Technische Universit\"at Dresden, Institute of Applied Physics, 01062 Dresden, Germany}

\author{Marisa A. Choffel}%
\affiliation{Department of Chemistry, Materials Science Institute, University of Oregon, Eugene, Oregon 97403, United States}

\author{Danielle M. Hamann}%
\affiliation{Department of Chemistry, Materials Science Institute, University of Oregon, Eugene, Oregon 97403, United States}

\author{David C. Johnson}%
\affiliation{Department of Chemistry, Materials Science Institute, University of Oregon, Eugene, Oregon 97403, United States}

\author{Kornelius Nielsch}%
\affiliation{Institute for Metallic Materials, Leibniz Institute of Solid State and Materials Science, 01069 Dresden, Germany}
\affiliation{Technische Universit\"at Dresden, Institute of Applied Physics, 01062 Dresden, Germany}
\affiliation{Technische Universit\"at Dresden, Institute of Materials Science, 01062 Dresden, Germany}%

\author{Andy Thomas}
\email[Corresponding author: ]{a.thomas@ifw-dresden.de}
\affiliation{Institute for Metallic Materials, Leibniz Institute of Solid State and Materials Science, 01069 Dresden, Germany}
\affiliation{Technische Universit\"at Dresden, Institut für Festk\"orper- und Materialphysik, 01062 Dresden, Germany}%
	
\date{\today}

\begin{abstract}
While a linear growth behavior is one of the fingerprints of textbook atomic layer deposition processes, the growth often deviates from that behavior in the initial regime, i.e. the first few cycles of a process. To properly understand the growth behavior in the initial regime is particularly important for applications that rely on the exact thickness of very thin films. The determination of the thicknesses of the initial regime, however, often requires special equipment and techniques that are not always available. We propose a thickness determination method that is based on X-ray reflectivity (XRR) measurements on double layer structures, i.e. substrate/base layer/top layer. XRR is a standard thin film characterization method. Utilizing the inherent properties of fast Fourier transformation in combination with a multi-Gaussian fitting routine permits the determination of thicknesses down to $t\approx\SI{2}{\nano\meter}$. We evaluate the boundaries of our model, which are given by the separation and full width at half maximum of the individual Gaussians. Finally, we compare our results with data from X-ray fluorescence spectroscopy, which is a standard method for measuring ultra thin films.
\end{abstract}

\maketitle

\begin{figure}
	\includegraphics[width=\columnwidth]{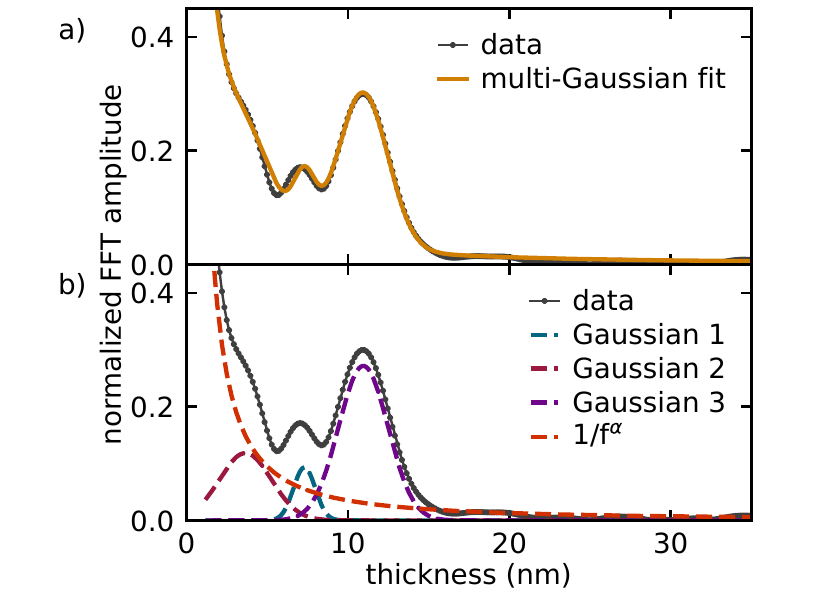}
	\caption{The FFT data as well as the multi-Gaussian fit to the full spectrum using Eq.\ref{eq:fitFunc} are shown in panel a). The fit is decomposed into its individual contributions in panel b). The maximum positions of Gaussian 1, 2 and 3 correspond to the thicknesses of layer 1, layer 2 and the sum of the layer thicknesses 1 and 2.}
	\label{fig:exp_plus_model}
\end{figure}

\begin{figure}
	\includegraphics[width=\columnwidth]{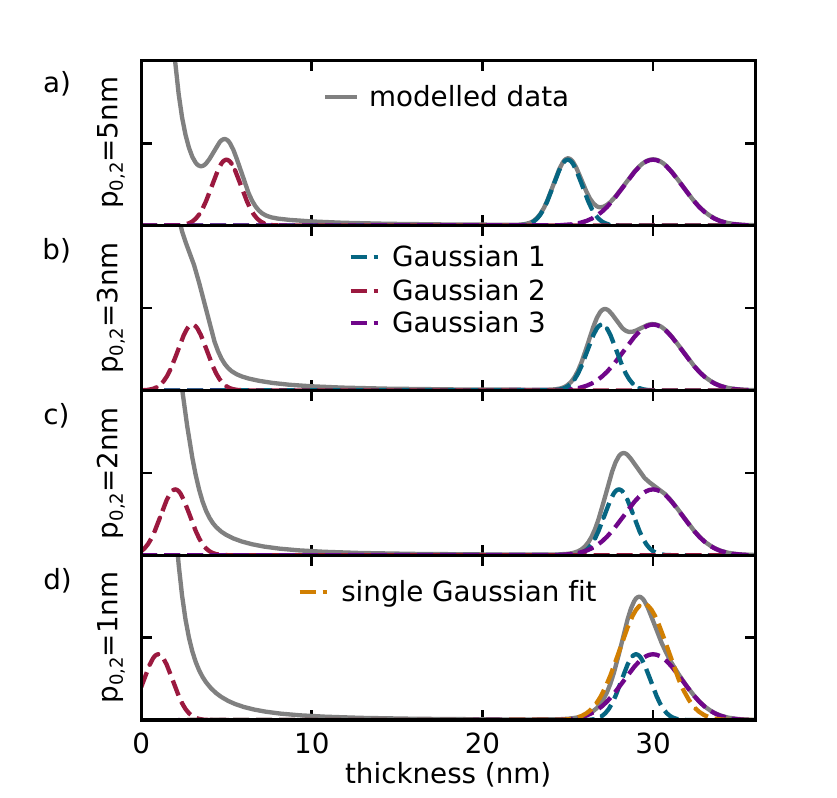}
	\caption{Panel a) to d) show modeled FFT data (grey curve) for two individual layer thicknesses (given by Gaussian 1 and 2) and their sum (given by Gaussian 3) including a 1/f$^\alpha$-background for different spacings of the maximum position $p_{0,2}$. Gaussian 1, 2 and 3 are given by the blue, red and purple dashed lines, respectively. A clear separation of Gaussian 3 (purple) from Gaussian 1 (blue) is possible down to a difference of 2 nm (panel c)). A single Gaussian fit to the peak in panel d) is given by the yellow dashed line emphasizing that there the resulting peak of the sum of Gaussian 3 and 1 can not be distinguished from one individual Gaussian anymore.}
	\label{fig:boundaries}
\end{figure}

Atomic layer deposition (ALD) is a key technology not only for the state-of-the-art semiconductor industry and nanoelectronics,\cite{Knez_2007,Kim_2009,Ritala_2009} but also for photovoltaics,\cite{Niu_2015,Delft_2012} catalysis\cite{Riha_2013,Schlicht_2018} and battery development.\cite{Meng_2012,Yu_2013} The two most characteristic features of ALD are the conformity and self-limitation of the growth process. One approach to test the self limiting behavior is by validating the linear increase of the layer thickness with the number of ALD cycles. However, it is common for ALD processes to deviate from this linear dependence for very low cycle numbers.\cite{Alam_2003,Lim_2001} Often a Volmer-Weber type growth is developed in the first few cycles, due to the different surface chemistries of the substrate and the layer itself.\cite{Venables_1984,Puurunen_2004} Understanding the initial growth stage is crucial for applications that rely on the exact thickness of ultra thin films, in particular, for the use in tunnel junctions, solar cells or as building blocks for nanolaminates.
The determination of low layer thicknesses within the initial stage is experimentally challenging and often involves the use of specific techniques and equipment, including X-ray fluorescence\cite{Dendooven_2011,Hamann_2018,Hung_2005} or various in-situ methods such as quarz crystal microbalance\cite{Fabreguette_2005,Wind_2010,Wiegand_2018} or vibrational sum-frequency.\cite{Vandalon_2019}
Herein, we propose an efficient way to determine the thickness of thin films down to \SI{2}{\nano\meter} using X-ray reflectivity and fast Fourier transformation (FFT) enabling the investigation of the initial stage for various ALD processes. The analysis utilizes double layer systems in combination with the inherent properties of a Fourier transformation to not only detect oscillation frequencies related to the single layers but also linear combinations of those oscillations. This simplifies the evaluation of XRR measurements by either fully eliminating the need of modeling the XRR data itself, which is often ambiguous due to the large number of involved parameters, or, in case a full modeling is indispensable, by providing a reasonable set of starting parameters. The proposed evaluation is well suited for of big data sets and enables the thickness determination with equipment, that is commonly available for many researchers in various fields. Additionally, it can be easily altered to describe any number of layers. We show the relevance of our approach as well as its limitation in the \si{\nano\meter}-regime. Furthermore, we compare our results for the substrate/\ch{Y2O3}/\ch{Fe2O3} layer system with X-ray fluorescence (XRF) data to validate our study.
\begin{figure*}[hbt!]
	\includegraphics[width=\linewidth]{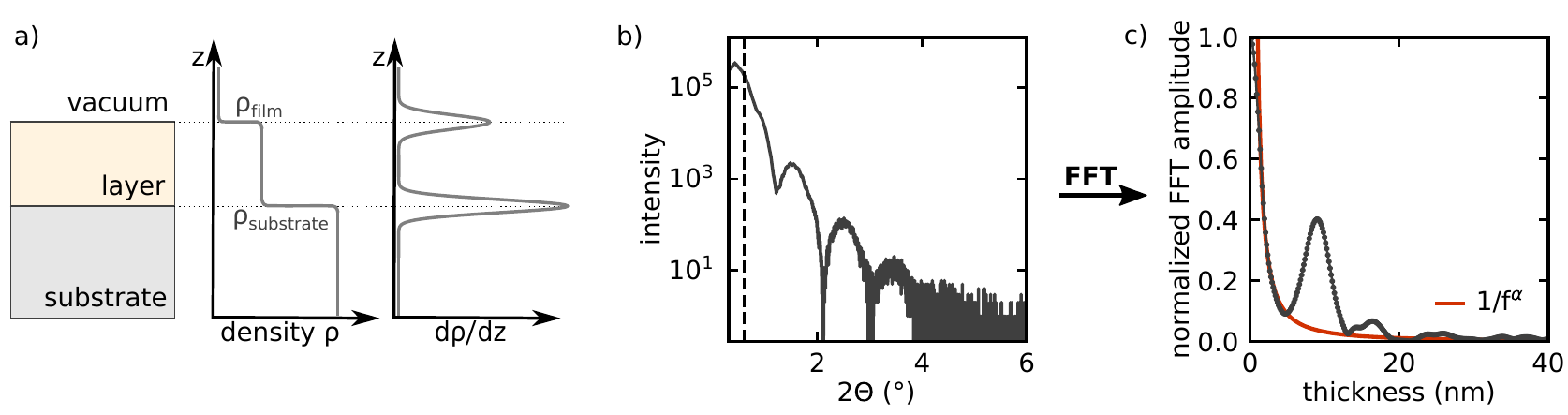}
	\caption{Panel a) depicts the schematic of a layer stack with one film on top of a substrate, the density function within the stack, as well as the change in density along the z- direction. Panel a) is adapted from Chason et al.\cite{Chason_1997} A X-ray reflectivity curve of a sample consisting of an ALD deposited \ch{Y_2O_3} layer on a \ch{Si}/\ch{SiO_2} substrate can be seen in panel b). The critical angle of total reflection is shown as the vertical dashed line. Applying a fast Fourier transform (FFT) to the windowed, zero padded data from panel b) leads to panel c). The 1/f$^{\alpha}$ noise in the low frequency range, which is given by the orange curve in panel c), is a characteristic of the FFT.}
	\label{fig:basics}
\end{figure*}
\\
X-ray reflectivity (XRR) is commonly employed to determine the thickness of a thin film or a layer system.\cite{Holy_1999,Daillant_2008,Chason_1997} This approach is based on the path difference of X-rays scattered by different interfaces of the  thin film sample. Figure \ref{fig:basics}a) shows a simple layer system: a substrate with one thin film on top. Under the assumption that each layer is a homogeneous medium with a mean refraction index smaller than unity, the classical Snell's law and the Fresnel equations can be applied to our system. The intensity of the reflected X-rays can then be expressed as a function of the Fourier transform of the electron density gradient normal to the surface. In the simplest case, the density change along the surface normal z at an interface of two materials is described by an error function (cp. Fig.\ref{fig:basics}a)). Differentiation of the error function leads to a Gaussian, \cite{Chason_1997} which describes the density gradient. For a stack with a single layer on a substrate this is depicted in Fig.\ref{fig:basics}a). A typical XRR measurement of one layer on a substrate, as is shown in Fig.\ref{fig:basics}b), exhibits oscillations in the reflected intensity which stem from the interference of X-rays reflected from the first and the second interface. The periodicity of these oscillations is proportional to the inverse of the layer thickness, whereas the amplitude of the oscillations depends on the density difference of the two adjacent layers. Hence, materials are best chosen with a density difference of at least 2\% for straightforward evaluation.\cite{Chason_1997} For data processing only incident angles larger than the critical angle of total reflection have to be taken into consideration. This critical angle was set to the point where the maximal intensity is reduced by a factor of 2, as is sketched by the dashed line in Fig.\ref{fig:basics}b).\\ 
Within the community of researchers experienced with X-ray techniques the concept of applying an FFT to XRR data to determine the thickness is well known and often used but is mostly discussed in terms of single layers or multilayer structures.\cite{Durand_2006,Sakurai_2008} Especially with the use of ALD spreading throughout different research fields due to the increasing number of available processes, we anticipate to enable researchers from other fields to perform this type of evaluation.
The fundamental characteristic of an FFT is to decompose a function into its frequency components. As discussed previously, XRR data consist of different oscillation frequencies which correspond to specific layer thicknesses. For a single layer applying an FFT to the autocorrelation function of the electronic density derivative along z (cp. Fig.\ref{fig:basics}a)) results in a spectrum which peaks at the thickness of the layer.\cite{Durand_2004} In Fig.\ref{fig:basics}c) this is illustrated for the data from panel b). To reduce spectral leakage a Hamming window was used on the data. Using the correct windowing function is especially crucial for the thickness determination of very thin films. After windowing the data was zero-padded to ten times its original length, which acts as an interpolation function on the FFT spectrum.\cite{Donelle_2005} The FFT was then performed on the windowed and padded data. Since the data sets are finite, the spectra have an 1/f$^\alpha$ background which is also sketched in Fig.\ref{fig:basics}c). For very thin single layers the corresponding thickness peak is concealed beneath this background, prohibiting the extraction of the layer thickness.\\
This issue can be solved by using systems with more than one layer. In the same fashion as for single layer systems, the FFT has been applied to multilayer systems.\cite{Sakurai_2008,Holy_1999} In the simplest case these consist of two deposited layers, i.e. substrate/base layer/top layer. For a two layer configuration, three peaks are expected in the FFT spectrum, which correspond to the thicknesses of the base layer, the top layer and the combined thickness of both layers. This characteristic can be used to determine the thickness of ultra thin layers for which the thickness extraction is usually not straightforward. As mentioned before, for single layers ($\mathrm{d}\leq\SI{5}{\nano\meter}$) the peak in the FFT spectrum is often concealed by the 1/f$^\alpha$ background. By using a two layer system, however, the thickness information of the thin layer is also contained within the combined thickness of both layers. Therefore, by evaluating the combined peak as well as the base layer peak the thickness of the (ultra) thin top layer can be extracted. The applicability of this evaluation method is independent of the explicit stacking order of the individual layers.\\
We applied the aforementioned evaluation method to determine the in-situ ALD growth of \ch{Fe2O3} on \ch{Y2O3} in the initial stage regime. Therefore, a series of samples featuring a constant \ch{Y2O3} base layer and different \ch{Fe2O3} top layers has been grown on \ch{Si}/\ch{SiO2} substrates with \SI{100}{\nano\meter} thermal oxide. The base layer always consists of 100 cycles of \ch{Y2O3}, which resulted in a mean thickness of $\SI[separate-uncertainty = true]{7.7(15)}{\nano\meter}$. Subsequently, different cycle numbers of \ch{Fe2O3} were deposited in-situ to form the top layer, resulting in a substrate/\ch{Y2O3}/\ch{Fe2O3} layer stack. For the XRR measurements a X'Pert diffractometer with a Cu anode from Philips Analytical was used. The initial beam path included a fixed incident beam mask (\SI{10}{\milli\meter}), a Soller slit (0.04 rad), a parabolic mirror for horizontal parallelization, a programmable Ni beam attenuator (\SI{0.125}{\milli\meter}) and a programmable divergence slit (1/16°). Finally, a 5° fixed horizontal slit is used to account for the sample size. The reflected beam is collimated by a parallel plate collimator (0.09°) before being collected by a PW1711 proportional detector. This configuration yields an initial intensity of $\approx 10^6$ counts per second. The error of the thicknesses determined by XRR was estimated to be the minimum thickness resolution of the FFT method. This was calculated by $\Delta d = 1/T_{\mathrm{XRR}}$, where $T_{\mathrm{XRR}}=\frac{2}{\lambda} \sqrt{\cos^2(\frac{2\theta_{c}}{2}\frac{2\pi}{180})-\cos^2(\frac{2\theta}{2}\frac{2\pi}{180})}$ is the length of the real space data with included autocorrelation relation. Therein, $\lambda = \SI{0.152}{\nano\meter}$ is the wavelength of the Cu-$K_\alpha$ radiation, $2\theta = \SI{6}{\degree}$  is the measurement range and the critical angle for both \ch{Fe2O3} and \ch{Y2O3} was determined to be $2\theta_c = \SI{0.6}{\degree}$. This relation results in a resolution of $\Delta d =\SI{1.5}{\nano\meter}$.\\
The FFT data of a typical sample (i.e. \ch{Y2O3} on \ch{Fe2O3}) can be found in Fig.\ref{fig:exp_plus_model}a). It shows two peaks as well as a shoulder in the 1/f$^\alpha$ background indicating a third peak. To extract the individual layer thicknesses from the experimental data, we used a multi-Gaussian fit, which is given by Eq.\ref{eq:fitFunc}. The fit function consists of the sum of three Gaussian functions plus the 1/f$^\alpha$ background. 
\begin{equation}
f_{\mathrm{fit}}(x)=\sum_{i=1}^{3}a_i\cdot\exp\left[-\ln2\cdot\left(\frac{p_{\mathrm{0,i}}-x}{w_i/2}\right)^2\right]+a_N\cdot\frac{1}{x^\alpha}
\label{eq:fitFunc}
\end{equation}
Here, $a_i$ describes the amplitudes of Gaussians 1, 2 and 3 and $a_N$ gives the amplitude of the 1/f$^\alpha$ noise. The maximum peak positions of the Gaussians are given by $p_{\mathrm{0}, i}$ and the full width at half maximum values by $w_i$. Additionally the relation $p_{\mathrm{0,2}}=p_{\mathrm{0,3}}-p_{\mathrm{0,1}}$ was introduced to account for the correlation of the maxima positions (i.e. the layer thicknesses).
The fit to the data in Fig.\ref{fig:exp_plus_model}a) using Eq.\ref{eq:fitFunc} is given by the yellow line. Panel b) shows the individual Gaussians remodeled with the parameters extracted by the fit. Although Gaussian 2 ($p_{\mathrm{0,2}}=\SI{3.7}{\nano\meter}$) is mostly concealed beneath the 1/f$^\alpha$ background (dashed orange line), it can still be resolved by our fitting procedure using Gaussian 3 ($p_{\mathrm{0,3}}=\SI{11.0}{\nano\meter}$) and Gaussian 1 ($p_{\mathrm{0,1}}=\SI{7.3}{\nano\meter}$), as is portrayed in Fig.\ref{fig:exp_plus_model}b). Please note that the thickness determination of a single layer with $d=\SI{3.7}{\nano\meter}(=p_{\mathrm{0,2}})$ would not be possible, since in that case the information would be hidden under the 1/f$^\alpha$ background. For using a double layer system, however, the thickness information of layer 2 is also contained in the peak of the sum of the layers enabling us to extract the thickness of layer 2 although it is concealed by the background. Comparing these thickness values determined by the multi-Gaussian fit with thicknesses determined from standard model fitting of the XRR data, a good accordance of the values can be observed. 
\\
To elucidate the boundaries of our multi-Gaussian approach we designed a model two layer system. Figure \ref{fig:boundaries} shows the modeled FFT spectrum consisting of two separate Gaussians (1 and 2) representing the thicknesses of the two individual layers, while Gaussian 3 represents the sum of the thicknesses of the individual layers. 
The maximum positions $p_{0,i}$ ($i=1,2,3$) of Gaussian 1, Gaussian 2, and Gaussian 3 define the layer thicknesses of layer 1, layer 2 and the sum of layers 1 and 2, respectively. The amplitudes and widths of the Gaussians were chosen to represent results of fitting the experimental data. The different panels show the change of the spectrum for several spacings of the maximum values of Gaussian 3 and Gaussian 1, which is $p_{0,2}$. For spacings $p_{0,2}\geq\SI{2}{nm}$, which are depicted in panels a) to c), Gaussian 1 and Gaussian 3 can be readily distinguished. A single Gaussian fit to the conjunction of Gaussian 1 and Gaussian 3 in the transient regime, i.e. panels b) and c), would lead to a substantial systematic error. For a separation of $p_{0,2}\leq\SI{1}{nm}$, however, the individual determination of both of the peaks is no longer unambiguously possible (cp. panel d). In this case representing the data by a single Gaussian fit is possible, in contrast to panels a) to c). Please be reminded, that the spacing between Gaussian 1 and 3 (i.e. $p_{0,2}$) describes the layer thickness of layer 2. Gaussian 2 is concealed beneath the 1/f$^\alpha$ background for $p_{0,2}\leq\SI{5}{nm}$, so the determination of the thickness of layer 2 is only indirectly possible by using Gaussian 1 and Gaussian 3.  Therefore, the minimum layer thickness that can be resolved with our setup using the multi-Gaussian fitting method lies between $\SIrange[range-phrase = -]{1}{2}{\nano\metre}$. Experimentally this lower boundary depends on the width of the individual Gaussians, which in turn depend on a collection of parameters arising from the specifics of the X-ray reflectometer used and the samples themselves.\cite{Balzar_1993,Sevenhans_1986,Delhez_1978} Please note that while any surface roughness of the interfaces does not affect the periodicity of the fringes per se, rough surfaces lead to a damping of the amplitude of the interference fringes as well as enhancing the general intensity decay.\cite{Sinha_1988,Sakurai_2008} This further reduces the possible resolution of the FFT by reducing the length of the data set. The thickness resolution can be increased by extending the measurement range as long as the signal remains above the noise floor, which could be achieved using a higher initial X-ray intensity. This can be realized by using a more powerful X-ray tube such as the $\SI{9}{\kilo\watt}$ rotating anode X-ray from Rigaku for which the maximum intensity was estimated to be $10^{7}$ cps. With this initial intensity a minimum resolution of $\Delta d=\SI{0.7}{\nano\meter}$ can be obtained. The evaluation script and example data are published on arXiv.\cite{code_DOI}
\\
We used the multi-Gaussian fitting routine to examine the layer stack series mentioned previously, i.e. substrate/\ch{Y2O3}/\ch{Fe2O3}. We compared our multi-Gaussian approach with the results of XRF measurements on the same samples.\cite{Hamann_2018} 
\begin{figure}[hbt!]
	\includegraphics{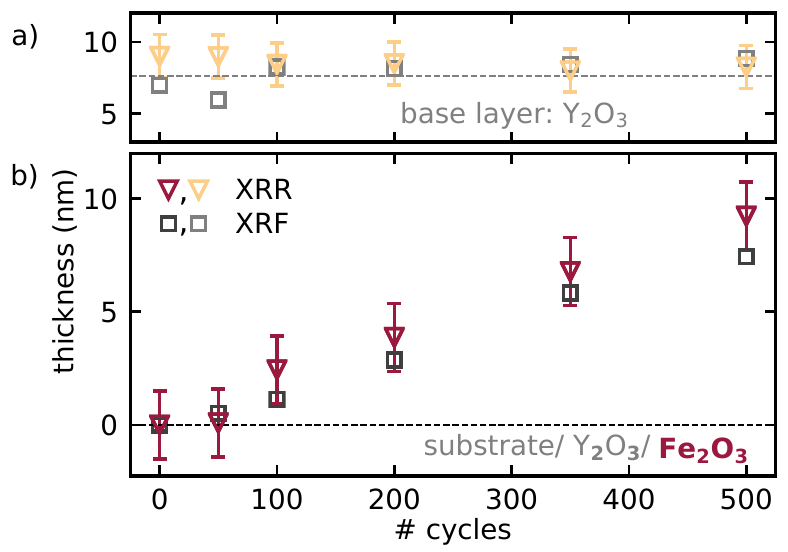}
	\caption{The base and top layer thicknesses in the ultra-low thickness regime of in-situ deposited \ch{Fe2O3} on substrate/\ch{Y2O3} are given in panels a) and b), respectively. The thicknesses determined by XRF measurements are shown as squares while the evaluation of the same samples by XRR utilizing the introduced multi-Gaussian fitting of the FFT spectra are depicted as triangles. The dashed gray line in panel a) gives the mean base layer thickness.}
	\label{fig:XRF}
\end{figure}
The thicknesses of the \ch{Y2O3} base layer extracted by both techniques are given in Fig.\ref{fig:XRF}a). The corresponding thicknesses of the \ch{Fe2O3} top layer are depicted in panel b). From the XRF measurements the number of atoms per area can be determined. A calibration by means of uniform films was done for the number of Fe atoms, whereas a drop casting method was used for the calibration of the Y atoms. The thicknesses of the layers were then calculated from the number of atoms per area using Eq.\ref{eq:thicknessXRF} by taking into account the volume of the unit cell ($V_{\mathrm{unit\,cell}}$) as well as the number of Fe or Y atoms per unit cell ($\#\;\mathrm{atoms}_{\mathrm{unit\,cell}}$). For the \ch{Fe2O3} layers a hexagonal unit cell with a volume of $V_{\mathrm{unit\,cell,Fe}} = \SI{0.9}{\nano\meter^3}$ and 12 atoms per unit cell was used.\cite{Wang_1998,Landolt-Boernstein_hematite} A cubic unit cell with $V_{\mathrm{unit\,cell, Y}} = \SI{1.2}{\nano\meter^3}$ and 11 atoms per unit cell was used for the \ch{Y2O3} layers.\cite{Landolt-Boernstein_yttria} The evaluated area of the XRF measurements was defined by the used aperture to be $A_{\mathrm{XRF}}=7.85\times10^{-5}\si{\square\meter}$.
\begin{equation}
t_{layer}= \frac{\#\;\mathrm{atoms_{XRF}}/A_{\mathrm{XRF}}}{\#\;\mathrm{atoms}_{\mathrm{unit\,cell}}/V_{\mathrm{unit\,cell}}}
\label{eq:thicknessXRF}
\end{equation}
A tooling factor of 1.37 [0.41] was determined by least square fitting for the \ch{Y2O3} [\ch{Fe2O3}] layer to achieve the best possible agreement between XRF and XRR values.\\
Both evaluation techniques result, within the margin of error, in the same layer thicknesses of top and base layer, corroborating the validity of the multi-Gaussian fitting routine for this layer system (cp. Fig\ref{fig:XRF}a) and b)). Both evaluation methods suggest a growth delay of the \ch{Fe2O3} top layer between 50 and 100 cycles, while the \ch{Y2O3} base layer stays constant, as expected. The evolution of the difference between the top layer thicknesses determined by XRR and XRF cannot be defined unambiguously. Within the experimental limits of the XRR evaluation, a description using a mean difference of \SI{0.8}{\nano\meter} is as justifiable as one utilizing a linear increase of the thickness difference with increasing cycle number. This linear dependence most likely results from an imperfect determination of the tooling factor, which leads to an increasing difference for increasing cycle numbers. Altogether, both the top and base layer thicknesses are in good agreement for the two different evaluation methods.

In summary, we described a complete but concise way of utilizing the inherent properties of the FFT to determine layer thicknesses down to \SI{2}{\nano\meter} from data acquired by XRR. Using bilayer structures consisting of substrate/base layer/top layer enables us to calculate the top layer thickness from the peak stemming from the combined thickness of base and top layer and the base layer peak. Using a Hamming window on the data as well as zero padding the data prior to the FFT is crucial to get the correct layer thicknesses, especially for very thin top layers. We describe the FFT spectrum using a multi-Gaussian fit. This approach is limited by the ability to distinguish between one Gaussian versus the superposition of two Gaussians, yielding a lowest layer thickness of \SI{2}{\nano\meter} for our samples and FWHM of the XRR peaks. We used the multi-Gaussian fitting approach to determine the initial stage of in-situ ALD grown of \ch{Fe2O3} layers on \ch{Y2O3}. For atomic layer deposition the determination of layer thicknesses within the initial regime is often crucial to guarantee very thin layers of exact thicknesses. Comparing the results to XRF data shows a good congruence of the layer thicknesses. Our approach is not limited to ALD
but can also be used for any kind of thin film deposition
technique. Furthermore, it enables the evaluation of in-situ as well as ex-situ grown samples, providing a wide
area of application. Using the presented evaluation could
help a wide variety of researchers to better understand
the initial stage of their (ALD) processes using customary equipment and a simple analysis procedure.\\

Authors to whom correspondence should be addressed: m.lammel@ifw-dresden.de, a.thomas@ifw-dresden.de

\section*{data availability}
Data available in article. Evaluation script as well as exemplary raw measurement data available on the arXiv (arXiv:2008.04626). Additional raw measurement data available on request from the authors.

\bibliography{references_dpp.bib}

\begin{thebibliography}{35}%
\makeatletter
\providecommand \@ifxundefined [1]{%
 \@ifx{#1\undefined}
}%
\providecommand \@ifnum [1]{%
 \ifnum #1\expandafter \@firstoftwo
 \else \expandafter \@secondoftwo
 \fi
}%
\providecommand \@ifx [1]{%
 \ifx #1\expandafter \@firstoftwo
 \else \expandafter \@secondoftwo
 \fi
}%
\providecommand \natexlab [1]{#1}%
\providecommand \enquote  [1]{``#1''}%
\providecommand \bibnamefont  [1]{#1}%
\providecommand \bibfnamefont [1]{#1}%
\providecommand \citenamefont [1]{#1}%
\providecommand \href@noop [0]{\@secondoftwo}%
\providecommand \href [0]{\begingroup \@sanitize@url \@href}%
\providecommand \@href[1]{\@@startlink{#1}\@@href}%
\providecommand \@@href[1]{\endgroup#1\@@endlink}%
\providecommand \@sanitize@url [0]{\catcode `\\12\catcode `\$12\catcode
  `\&12\catcode `\#12\catcode `\^12\catcode `\_12\catcode `\%12\relax}%
\providecommand \@@startlink[1]{}%
\providecommand \@@endlink[0]{}%
\providecommand \url  [0]{\begingroup\@sanitize@url \@url }%
\providecommand \@url [1]{\endgroup\@href {#1}{\urlprefix }}%
\providecommand \urlprefix  [0]{URL }%
\providecommand \Eprint [0]{\href }%
\providecommand \doibase [0]{http://dx.doi.org/}%
\providecommand \selectlanguage [0]{\@gobble}%
\providecommand \bibinfo  [0]{\@secondoftwo}%
\providecommand \bibfield  [0]{\@secondoftwo}%
\providecommand \translation [1]{[#1]}%
\providecommand \BibitemOpen [0]{}%
\providecommand \bibitemStop [0]{}%
\providecommand \bibitemNoStop [0]{.\EOS\space}%
\providecommand \EOS [0]{\spacefactor3000\relax}%
\providecommand \BibitemShut  [1]{\csname bibitem#1\endcsname}%
\let\auto@bib@innerbib\@empty
\bibitem [{\citenamefont {Knez}, \citenamefont {Nielsch},\ and\ \citenamefont
  {Niinistö}(2007)}]{Knez_2007}%
  \BibitemOpen
  \bibfield  {author} {\bibinfo {author} {\bibfnamefont {M.}~\bibnamefont
  {Knez}}, \bibinfo {author} {\bibfnamefont {K.}~\bibnamefont {Nielsch}}, \
  and\ \bibinfo {author} {\bibfnamefont {L.}~\bibnamefont {Niinistö}},\
  }\bibfield  {title} {\enquote {\bibinfo {title} {Synthesis and surface
  engineering of complex nanostructures by atomic layer deposition},}\ }\href
  {\doibase 10.1002/adma.200700079} {\bibfield  {journal} {\bibinfo  {journal}
  {Advanced Materials}\ }\textbf {\bibinfo {volume} {19}},\ \bibinfo {pages}
  {3425--3438} (\bibinfo {year} {2007})},\ \Eprint
  {http://arxiv.org/abs/https://onlinelibrary.wiley.com/doi/pdf/10.1002/adma.200700079}
  {https://onlinelibrary.wiley.com/doi/pdf/10.1002/adma.200700079} \BibitemShut
  {NoStop}%
\bibitem [{\citenamefont {Kim}, \citenamefont {Lee},\ and\ \citenamefont
  {Maeng}(2009)}]{Kim_2009}%
  \BibitemOpen
  \bibfield  {author} {\bibinfo {author} {\bibfnamefont {H.}~\bibnamefont
  {Kim}}, \bibinfo {author} {\bibfnamefont {H.-B.-R.}\ \bibnamefont {Lee}}, \
  and\ \bibinfo {author} {\bibfnamefont {W.-J.}\ \bibnamefont {Maeng}},\
  }\bibfield  {title} {\enquote {\bibinfo {title} {Applications of atomic layer
  deposition to nanofabrication and emerging nanodevices},}\ }\href {\doibase
  https://doi.org/10.1016/j.tsf.2008.09.007} {\bibfield  {journal} {\bibinfo
  {journal} {Thin Solid Films}\ }\textbf {\bibinfo {volume} {517}},\ \bibinfo
  {pages} {2563 -- 2580} (\bibinfo {year} {2009})}\BibitemShut {NoStop}%
\bibitem [{\citenamefont {Ritala}\ and\ \citenamefont
  {Niinist\"o}(2009)}]{Ritala_2009}%
  \BibitemOpen
  \bibfield  {author} {\bibinfo {author} {\bibfnamefont {M.}~\bibnamefont
  {Ritala}}\ and\ \bibinfo {author} {\bibfnamefont {J.}~\bibnamefont
  {Niinist\"o}},\ }\bibfield  {title} {\enquote {\bibinfo {title} {Industrial
  applications of atomic layer deposition},}\ }\bibfield  {booktitle} {\emph
  {\bibinfo {booktitle} {{ECS} Transactions}},\ }\href {\doibase
  10.1149/1.3207651} {\  (\bibinfo {year} {2009}),\
  10.1149/1.3207651}\BibitemShut {NoStop}%
\bibitem [{\citenamefont {Niu}\ \emph {et~al.}(2015)\citenamefont {Niu},
  \citenamefont {Li}, \citenamefont {Karuturi}, \citenamefont {Fam},
  \citenamefont {Fan}, \citenamefont {Shrestha}, \citenamefont {Wong},\ and\
  \citenamefont {Tok}}]{Niu_2015}%
  \BibitemOpen
  \bibfield  {author} {\bibinfo {author} {\bibfnamefont {W.}~\bibnamefont
  {Niu}}, \bibinfo {author} {\bibfnamefont {X.}~\bibnamefont {Li}}, \bibinfo
  {author} {\bibfnamefont {S.~K.}\ \bibnamefont {Karuturi}}, \bibinfo {author}
  {\bibfnamefont {D.~W.}\ \bibnamefont {Fam}}, \bibinfo {author} {\bibfnamefont
  {H.}~\bibnamefont {Fan}}, \bibinfo {author} {\bibfnamefont {S.}~\bibnamefont
  {Shrestha}}, \bibinfo {author} {\bibfnamefont {L.~H.}\ \bibnamefont {Wong}},
  \ and\ \bibinfo {author} {\bibfnamefont {A.~I.~Y.}\ \bibnamefont {Tok}},\
  }\bibfield  {title} {\enquote {\bibinfo {title} {Applications of atomic layer
  deposition in solar cells},}\ }\href {\doibase 10.1088/0957-4484/26/6/064001}
  {\bibfield  {journal} {\bibinfo  {journal} {Nanotechnology}\ }\textbf
  {\bibinfo {volume} {26}},\ \bibinfo {pages} {064001} (\bibinfo {year}
  {2015})}\BibitemShut {NoStop}%
\bibitem [{\citenamefont {Delft}, \citenamefont {Garcia-Alonso},\ and\
  \citenamefont {Kessels}(2012)}]{Delft_2012}%
  \BibitemOpen
  \bibfield  {author} {\bibinfo {author} {\bibfnamefont {J.}~\bibnamefont
  {Delft}}, \bibinfo {author} {\bibfnamefont {D.}~\bibnamefont
  {Garcia-Alonso}}, \ and\ \bibinfo {author} {\bibfnamefont {W.}~\bibnamefont
  {Kessels}},\ }\bibfield  {title} {\enquote {\bibinfo {title} {Atomic layer
  deposition for photovoltaics: Applications and prospects for solar cell
  manufacturing},}\ }\href {\doibase 10.1088/0268-1242/27/7/074002} {\bibfield
  {journal} {\bibinfo  {journal} {Semiconductor Science and Technology}\
  }\textbf {\bibinfo {volume} {27}},\ \bibinfo {pages} {074002} (\bibinfo
  {year} {2012})}\BibitemShut {NoStop}%
\bibitem [{\citenamefont {Riha}\ \emph {et~al.}(2013)\citenamefont {Riha},
  \citenamefont {Klahr}, \citenamefont {Tyo}, \citenamefont {Seifert},
  \citenamefont {Vajda}, \citenamefont {Pellin}, \citenamefont {Hamann},\ and\
  \citenamefont {Martinson}}]{Riha_2013}%
  \BibitemOpen
  \bibfield  {author} {\bibinfo {author} {\bibfnamefont {S.~C.}\ \bibnamefont
  {Riha}}, \bibinfo {author} {\bibfnamefont {B.~M.}\ \bibnamefont {Klahr}},
  \bibinfo {author} {\bibfnamefont {E.~C.}\ \bibnamefont {Tyo}}, \bibinfo
  {author} {\bibfnamefont {S.}~\bibnamefont {Seifert}}, \bibinfo {author}
  {\bibfnamefont {S.}~\bibnamefont {Vajda}}, \bibinfo {author} {\bibfnamefont
  {M.~J.}\ \bibnamefont {Pellin}}, \bibinfo {author} {\bibfnamefont {T.~W.}\
  \bibnamefont {Hamann}}, \ and\ \bibinfo {author} {\bibfnamefont {A.~B.~F.}\
  \bibnamefont {Martinson}},\ }\bibfield  {title} {\enquote {\bibinfo {title}
  {Atomic layer deposition of a submonolayer catalyst for the enhanced
  photoelectrochemical performance of water oxidation with hematite},}\ }\href
  {\doibase 10.1021/nn305639z} {\bibfield  {journal} {\bibinfo  {journal} {ACS
  Nano}\ }\textbf {\bibinfo {volume} {7}},\ \bibinfo {pages} {2396--2405}
  (\bibinfo {year} {2013})},\ \bibinfo {note} {pMID: 23398051},\ \Eprint
  {http://arxiv.org/abs/https://doi.org/10.1021/nn305639z}
  {https://doi.org/10.1021/nn305639z} \BibitemShut {NoStop}%
\bibitem [{\citenamefont {Schlicht}\ \emph {et~al.}(2018)\citenamefont
  {Schlicht}, \citenamefont {Haschke}, \citenamefont {Mikhailovskii},
  \citenamefont {Manshina},\ and\ \citenamefont {Bachmann}}]{Schlicht_2018}%
  \BibitemOpen
  \bibfield  {author} {\bibinfo {author} {\bibfnamefont {S.}~\bibnamefont
  {Schlicht}}, \bibinfo {author} {\bibfnamefont {S.}~\bibnamefont {Haschke}},
  \bibinfo {author} {\bibfnamefont {V.}~\bibnamefont {Mikhailovskii}}, \bibinfo
  {author} {\bibfnamefont {A.}~\bibnamefont {Manshina}}, \ and\ \bibinfo
  {author} {\bibfnamefont {J.}~\bibnamefont {Bachmann}},\ }\bibfield  {title}
  {\enquote {\bibinfo {title} {Highly reversible water oxidation at ordered
  nanoporous iridium electrodes based on an original atomic layer
  deposition},}\ }\href {\doibase 10.1002/celc.201800152} {\bibfield  {journal}
  {\bibinfo  {journal} {ChemElectroChem}\ }\textbf {\bibinfo {volume} {5}},\
  \bibinfo {pages} {1259--1264} (\bibinfo {year} {2018})},\ \Eprint
  {http://arxiv.org/abs/https://chemistry-europe.onlinelibrary.wiley.com/doi/pdf/10.1002/celc.201800152}
  {https://chemistry-europe.onlinelibrary.wiley.com/doi/pdf/10.1002/celc.201800152}
  \BibitemShut {NoStop}%
\bibitem [{\citenamefont {Meng}, \citenamefont {Yang},\ and\ \citenamefont
  {Sun}(2012)}]{Meng_2012}%
  \BibitemOpen
  \bibfield  {author} {\bibinfo {author} {\bibfnamefont {X.}~\bibnamefont
  {Meng}}, \bibinfo {author} {\bibfnamefont {X.-Q.}\ \bibnamefont {Yang}}, \
  and\ \bibinfo {author} {\bibfnamefont {X.}~\bibnamefont {Sun}},\ }\bibfield
  {title} {\enquote {\bibinfo {title} {Emerging applications of atomic layer
  deposition for lithium-ion battery studies},}\ }\href {\doibase
  10.1002/adma.201200397} {\bibfield  {journal} {\bibinfo  {journal} {Advanced
  Materials}\ }\textbf {\bibinfo {volume} {24}},\ \bibinfo {pages} {3589--3615}
  (\bibinfo {year} {2012})},\ \Eprint
  {http://arxiv.org/abs/https://onlinelibrary.wiley.com/doi/pdf/10.1002/adma.201200397}
  {https://onlinelibrary.wiley.com/doi/pdf/10.1002/adma.201200397} \BibitemShut
  {NoStop}%
\bibitem [{\citenamefont {Lei}\ \emph {et~al.}(2013)\citenamefont {Lei},
  \citenamefont {Lu}, \citenamefont {Luo}, \citenamefont {Wu}, \citenamefont
  {Du}, \citenamefont {Zhang}, \citenamefont {Ren}, \citenamefont {Wen},
  \citenamefont {Miller}, \citenamefont {Miller}, \citenamefont {Sun},
  \citenamefont {Elam},\ and\ \citenamefont {Amine}}]{Yu_2013}%
  \BibitemOpen
  \bibfield  {author} {\bibinfo {author} {\bibfnamefont {Y.}~\bibnamefont
  {Lei}}, \bibinfo {author} {\bibfnamefont {J.}~\bibnamefont {Lu}}, \bibinfo
  {author} {\bibfnamefont {X.}~\bibnamefont {Luo}}, \bibinfo {author}
  {\bibfnamefont {T.}~\bibnamefont {Wu}}, \bibinfo {author} {\bibfnamefont
  {P.}~\bibnamefont {Du}}, \bibinfo {author} {\bibfnamefont {X.}~\bibnamefont
  {Zhang}}, \bibinfo {author} {\bibfnamefont {Y.}~\bibnamefont {Ren}}, \bibinfo
  {author} {\bibfnamefont {J.}~\bibnamefont {Wen}}, \bibinfo {author}
  {\bibfnamefont {D.~J.}\ \bibnamefont {Miller}}, \bibinfo {author}
  {\bibfnamefont {J.~T.}\ \bibnamefont {Miller}}, \bibinfo {author}
  {\bibfnamefont {Y.-K.}\ \bibnamefont {Sun}}, \bibinfo {author} {\bibfnamefont
  {J.~W.}\ \bibnamefont {Elam}}, \ and\ \bibinfo {author} {\bibfnamefont
  {K.}~\bibnamefont {Amine}},\ }\bibfield  {title} {\enquote {\bibinfo {title}
  {Synthesis of porous carbon supported palladium nanoparticle catalysts by
  atomic layer deposition: Application for rechargeable lithium–o2
  battery},}\ }\href {\doibase 10.1021/nl401833p} {\bibfield  {journal}
  {\bibinfo  {journal} {Nano Letters}\ }\textbf {\bibinfo {volume} {13}},\
  \bibinfo {pages} {4182--4189} (\bibinfo {year} {2013})},\ \bibinfo {note}
  {pMID: 23927754},\ \Eprint
  {http://arxiv.org/abs/https://doi.org/10.1021/nl401833p}
  {https://doi.org/10.1021/nl401833p} \BibitemShut {NoStop}%
\bibitem [{\citenamefont {Alam}\ and\ \citenamefont {Green}(2003)}]{Alam_2003}%
  \BibitemOpen
  \bibfield  {author} {\bibinfo {author} {\bibfnamefont {M.~A.}\ \bibnamefont
  {Alam}}\ and\ \bibinfo {author} {\bibfnamefont {M.~L.}\ \bibnamefont
  {Green}},\ }\bibfield  {title} {\enquote {\bibinfo {title} {Mathematical
  description of atomic layer deposition and its application to the nucleation
  and growth of hfo2 gate dielectric layers},}\ }\href {\doibase
  10.1063/1.1599978} {\bibfield  {journal} {\bibinfo  {journal} {Journal of
  Applied Physics}\ }\textbf {\bibinfo {volume} {94}},\ \bibinfo {pages}
  {3403--3413} (\bibinfo {year} {2003})},\ \Eprint
  {http://arxiv.org/abs/https://doi.org/10.1063/1.1599978}
  {https://doi.org/10.1063/1.1599978} \BibitemShut {NoStop}%
\bibitem [{\citenamefont {Lim}, \citenamefont {Park},\ and\ \citenamefont
  {Kang}(2001)}]{Lim_2001}%
  \BibitemOpen
  \bibfield  {author} {\bibinfo {author} {\bibfnamefont {J.-W.}\ \bibnamefont
  {Lim}}, \bibinfo {author} {\bibfnamefont {H.-S.}\ \bibnamefont {Park}}, \
  and\ \bibinfo {author} {\bibfnamefont {S.-W.}\ \bibnamefont {Kang}},\
  }\bibfield  {title} {\enquote {\bibinfo {title} {Kinetic modeling of film
  growth rate in atomic layer deposition},}\ }\href {\doibase
  10.1149/1.1368102} {\bibfield  {journal} {\bibinfo  {journal} {Journal of The
  Electrochemical Society}\ }\textbf {\bibinfo {volume} {148}},\ \bibinfo
  {pages} {C403} (\bibinfo {year} {2001})}\BibitemShut {NoStop}%
\bibitem [{\citenamefont {Venables}, \citenamefont {Spiller},\ and\
  \citenamefont {Hanbucken}(1984)}]{Venables_1984}%
  \BibitemOpen
  \bibfield  {author} {\bibinfo {author} {\bibfnamefont {J.~A.}\ \bibnamefont
  {Venables}}, \bibinfo {author} {\bibfnamefont {G.~D.~T.}\ \bibnamefont
  {Spiller}}, \ and\ \bibinfo {author} {\bibfnamefont {M.}~\bibnamefont
  {Hanbucken}},\ }\bibfield  {title} {\enquote {\bibinfo {title} {Nucleation
  and growth of thin films},}\ }\href {\doibase 10.1088/0034-4885/47/4/002}
  {\bibfield  {journal} {\bibinfo  {journal} {Reports on Progress in Physics}\
  }\textbf {\bibinfo {volume} {47}},\ \bibinfo {pages} {399--459} (\bibinfo
  {year} {1984})}\BibitemShut {NoStop}%
\bibitem [{\citenamefont {Puurunen}\ and\ \citenamefont
  {Vandervorst}(2004)}]{Puurunen_2004}%
  \BibitemOpen
  \bibfield  {author} {\bibinfo {author} {\bibfnamefont {R.~L.}\ \bibnamefont
  {Puurunen}}\ and\ \bibinfo {author} {\bibfnamefont {W.}~\bibnamefont
  {Vandervorst}},\ }\bibfield  {title} {\enquote {\bibinfo {title} {Island
  growth as a growth mode in atomic layer deposition: A phenomenological
  model},}\ }\href {\doibase 10.1063/1.1810193} {\bibfield  {journal} {\bibinfo
   {journal} {Journal of Applied Physics}\ }\textbf {\bibinfo {volume} {96}},\
  \bibinfo {pages} {7686--7695} (\bibinfo {year} {2004})},\ \Eprint
  {http://arxiv.org/abs/https://doi.org/10.1063/1.1810193}
  {https://doi.org/10.1063/1.1810193} \BibitemShut {NoStop}%
\bibitem [{\citenamefont {Dendooven}\ \emph {et~al.}(2011)\citenamefont
  {Dendooven}, \citenamefont {Pulinthanathu~Sree}, \citenamefont {De~Keyser},
  \citenamefont {Deduytsche}, \citenamefont {Martens}, \citenamefont {Ludwig},\
  and\ \citenamefont {Detavernier}}]{Dendooven_2011}%
  \BibitemOpen
  \bibfield  {author} {\bibinfo {author} {\bibfnamefont {J.}~\bibnamefont
  {Dendooven}}, \bibinfo {author} {\bibfnamefont {S.}~\bibnamefont
  {Pulinthanathu~Sree}}, \bibinfo {author} {\bibfnamefont {K.}~\bibnamefont
  {De~Keyser}}, \bibinfo {author} {\bibfnamefont {D.}~\bibnamefont
  {Deduytsche}}, \bibinfo {author} {\bibfnamefont {J.~A.}\ \bibnamefont
  {Martens}}, \bibinfo {author} {\bibfnamefont {K.~F.}\ \bibnamefont {Ludwig}},
  \ and\ \bibinfo {author} {\bibfnamefont {C.}~\bibnamefont {Detavernier}},\
  }\bibfield  {title} {\enquote {\bibinfo {title} {In situ x-ray fluorescence
  measurements during atomic layer deposition: Nucleation and growth of tio2 on
  planar substrates and in nanoporous films},}\ }\href {\doibase
  10.1021/jp111314b} {\bibfield  {journal} {\bibinfo  {journal} {The Journal of
  Physical Chemistry C}\ }\textbf {\bibinfo {volume} {115}},\ \bibinfo {pages}
  {6605--6610} (\bibinfo {year} {2011})},\ \Eprint
  {http://arxiv.org/abs/https://doi.org/10.1021/jp111314b}
  {https://doi.org/10.1021/jp111314b} \BibitemShut {NoStop}%
\bibitem [{\citenamefont {Hamann}\ \emph {et~al.}(2018)\citenamefont {Hamann},
  \citenamefont {Bardgett}, \citenamefont {Cordova}, \citenamefont {Maynard},
  \citenamefont {Hadland}, \citenamefont {Lygo}, \citenamefont {Wood},
  \citenamefont {Esters},\ and\ \citenamefont {Johnson}}]{Hamann_2018}%
  \BibitemOpen
  \bibfield  {author} {\bibinfo {author} {\bibfnamefont {D.~M.}\ \bibnamefont
  {Hamann}}, \bibinfo {author} {\bibfnamefont {D.}~\bibnamefont {Bardgett}},
  \bibinfo {author} {\bibfnamefont {D.~L.~M.}\ \bibnamefont {Cordova}},
  \bibinfo {author} {\bibfnamefont {L.~A.}\ \bibnamefont {Maynard}}, \bibinfo
  {author} {\bibfnamefont {E.~C.}\ \bibnamefont {Hadland}}, \bibinfo {author}
  {\bibfnamefont {A.~C.}\ \bibnamefont {Lygo}}, \bibinfo {author}
  {\bibfnamefont {S.~R.}\ \bibnamefont {Wood}}, \bibinfo {author}
  {\bibfnamefont {M.}~\bibnamefont {Esters}}, \ and\ \bibinfo {author}
  {\bibfnamefont {D.~C.}\ \bibnamefont {Johnson}},\ }\bibfield  {title}
  {\enquote {\bibinfo {title} {Sub-monolayer accuracy in determining the number
  of atoms per unit area in ultrathin films using x-ray fluorescence},}\ }\href
  {\doibase 10.1021/acs.chemmater.8b02591} {\bibfield  {journal} {\bibinfo
  {journal} {Chemistry of Materials}\ }\textbf {\bibinfo {volume} {30}},\
  \bibinfo {pages} {6209--6216} (\bibinfo {year} {2018})},\ \Eprint
  {http://arxiv.org/abs/https://doi.org/10.1021/acs.chemmater.8b02591}
  {https://doi.org/10.1021/acs.chemmater.8b02591} \BibitemShut {NoStop}%
\bibitem [{\citenamefont {Hung}\ \emph {et~al.}(2005)\citenamefont {Hung},
  \citenamefont {Gondran}, \citenamefont {Ghatak-Roy}, \citenamefont {Terada},
  \citenamefont {Bunday}, \citenamefont {Yeung},\ and\ \citenamefont
  {Diebold}}]{Hung_2005}%
  \BibitemOpen
  \bibfield  {author} {\bibinfo {author} {\bibfnamefont {P.~Y.}\ \bibnamefont
  {Hung}}, \bibinfo {author} {\bibfnamefont {C.}~\bibnamefont {Gondran}},
  \bibinfo {author} {\bibfnamefont {A.}~\bibnamefont {Ghatak-Roy}}, \bibinfo
  {author} {\bibfnamefont {S.}~\bibnamefont {Terada}}, \bibinfo {author}
  {\bibfnamefont {B.}~\bibnamefont {Bunday}}, \bibinfo {author} {\bibfnamefont
  {H.}~\bibnamefont {Yeung}}, \ and\ \bibinfo {author} {\bibfnamefont
  {A.}~\bibnamefont {Diebold}},\ }\bibfield  {title} {\enquote {\bibinfo
  {title} {X-ray reflectometry and x-ray fluorescence monitoring of the atomic
  layer deposition process for high-k gate dielectrics},}\ }\href {\doibase
  10.1116/1.2009774} {\bibfield  {journal} {\bibinfo  {journal} {Journal of
  Vacuum Science \& Technology B: Microelectronics and Nanometer Structures
  Processing, Measurement, and Phenomena}\ }\textbf {\bibinfo {volume} {23}},\
  \bibinfo {pages} {2244--2248} (\bibinfo {year} {2005})},\ \Eprint
  {http://arxiv.org/abs/https://avs.scitation.org/doi/pdf/10.1116/1.2009774}
  {https://avs.scitation.org/doi/pdf/10.1116/1.2009774} \BibitemShut {NoStop}%
\bibitem [{\citenamefont {Fabreguette}\ \emph {et~al.}(2005)\citenamefont
  {Fabreguette}, \citenamefont {Sechrist}, \citenamefont {Elam},\ and\
  \citenamefont {George}}]{Fabreguette_2005}%
  \BibitemOpen
  \bibfield  {author} {\bibinfo {author} {\bibfnamefont {F.}~\bibnamefont
  {Fabreguette}}, \bibinfo {author} {\bibfnamefont {Z.}~\bibnamefont
  {Sechrist}}, \bibinfo {author} {\bibfnamefont {J.}~\bibnamefont {Elam}}, \
  and\ \bibinfo {author} {\bibfnamefont {S.}~\bibnamefont {George}},\
  }\bibfield  {title} {\enquote {\bibinfo {title} {Quartz crystal microbalance
  study of tungsten atomic layer deposition using wf6 and si2h6},}\ }\href
  {\doibase https://doi.org/10.1016/j.tsf.2005.04.114} {\bibfield  {journal}
  {\bibinfo  {journal} {Thin Solid Films}\ }\textbf {\bibinfo {volume} {488}},\
  \bibinfo {pages} {103 -- 110} (\bibinfo {year} {2005})}\BibitemShut {NoStop}%
\bibitem [{\citenamefont {Wind}\ and\ \citenamefont
  {George}(2010)}]{Wind_2010}%
  \BibitemOpen
  \bibfield  {author} {\bibinfo {author} {\bibfnamefont {R.~A.}\ \bibnamefont
  {Wind}}\ and\ \bibinfo {author} {\bibfnamefont {S.~M.}\ \bibnamefont
  {George}},\ }\bibfield  {title} {\enquote {\bibinfo {title} {Quartz crystal
  microbalance studies of al2o3 atomic layer deposition using trimethylaluminum
  and water at 125 °c},}\ }\href {\doibase 10.1021/jp9049268} {\bibfield
  {journal} {\bibinfo  {journal} {The Journal of Physical Chemistry A}\
  }\textbf {\bibinfo {volume} {114}},\ \bibinfo {pages} {1281--1289} (\bibinfo
  {year} {2010})},\ \bibinfo {note} {pMID: 19757806},\ \Eprint
  {http://arxiv.org/abs/https://doi.org/10.1021/jp9049268}
  {https://doi.org/10.1021/jp9049268} \BibitemShut {NoStop}%
\bibitem [{\citenamefont {Wiegand}\ \emph {et~al.}(2018)\citenamefont
  {Wiegand}, \citenamefont {Faust}, \citenamefont {Meinhardt}, \citenamefont
  {Blick}, \citenamefont {Zierold},\ and\ \citenamefont
  {Nielsch}}]{Wiegand_2018}%
  \BibitemOpen
  \bibfield  {author} {\bibinfo {author} {\bibfnamefont {C.~W.}\ \bibnamefont
  {Wiegand}}, \bibinfo {author} {\bibfnamefont {R.}~\bibnamefont {Faust}},
  \bibinfo {author} {\bibfnamefont {A.}~\bibnamefont {Meinhardt}}, \bibinfo
  {author} {\bibfnamefont {R.~H.}\ \bibnamefont {Blick}}, \bibinfo {author}
  {\bibfnamefont {R.}~\bibnamefont {Zierold}}, \ and\ \bibinfo {author}
  {\bibfnamefont {K.}~\bibnamefont {Nielsch}},\ }\bibfield  {title} {\enquote
  {\bibinfo {title} {Understanding the growth mechanisms of multilayered
  systems in atomic layer deposition process},}\ }\href {\doibase
  10.1021/acs.chemmater.7b05128} {\bibfield  {journal} {\bibinfo  {journal}
  {Chemistry of Materials}\ }\textbf {\bibinfo {volume} {30}},\ \bibinfo
  {pages} {1971--1979} (\bibinfo {year} {2018})},\ \Eprint
  {http://arxiv.org/abs/https://doi.org/10.1021/acs.chemmater.7b05128}
  {https://doi.org/10.1021/acs.chemmater.7b05128} \BibitemShut {NoStop}%
\bibitem [{\citenamefont {Vandalon}\ and\ \citenamefont
  {Kessels}(2019)}]{Vandalon_2019}%
  \BibitemOpen
  \bibfield  {author} {\bibinfo {author} {\bibfnamefont {V.}~\bibnamefont
  {Vandalon}}\ and\ \bibinfo {author} {\bibfnamefont {W.~M. M.~E.}\
  \bibnamefont {Kessels}},\ }\bibfield  {title} {\enquote {\bibinfo {title}
  {Initial growth study of atomic-layer deposition of al2o3 by vibrational
  sum-frequency generation},}\ }\href {\doibase 10.1021/acs.langmuir.9b01600}
  {\bibfield  {journal} {\bibinfo  {journal} {Langmuir}\ }\textbf {\bibinfo
  {volume} {35}},\ \bibinfo {pages} {10374--10382} (\bibinfo {year} {2019})},\
  \bibinfo {note} {pMID: 31310143},\ \Eprint
  {http://arxiv.org/abs/https://doi.org/10.1021/acs.langmuir.9b01600}
  {https://doi.org/10.1021/acs.langmuir.9b01600} \BibitemShut {NoStop}%
\bibitem [{\citenamefont {Chason}\ and\ \citenamefont
  {Mayer}(1997)}]{Chason_1997}%
  \BibitemOpen
  \bibfield  {author} {\bibinfo {author} {\bibfnamefont {E.}~\bibnamefont
  {Chason}}\ and\ \bibinfo {author} {\bibfnamefont {T.~M.}\ \bibnamefont
  {Mayer}},\ }\bibfield  {title} {\enquote {\bibinfo {title} {Thin film and
  surface characterization by specular x-ray reflectivity},}\ }\href {\doibase
  10.1080/10408439708241258} {\bibfield  {journal} {\bibinfo  {journal}
  {Critical Reviews in Solid State and Materials Sciences}\ }\textbf {\bibinfo
  {volume} {22}},\ \bibinfo {pages} {1--67} (\bibinfo {year} {1997})},\ \Eprint
  {http://arxiv.org/abs/https://doi.org/10.1080/10408439708241258}
  {https://doi.org/10.1080/10408439708241258} \BibitemShut {NoStop}%
\bibitem [{\citenamefont {Holy}, \citenamefont {Pietsch},\ and\ \citenamefont
  {Baumbach}(1999)}]{Holy_1999}%
  \BibitemOpen
  \bibfield  {author} {\bibinfo {author} {\bibfnamefont {V.}~\bibnamefont
  {Holy}}, \bibinfo {author} {\bibfnamefont {U.}~\bibnamefont {Pietsch}}, \
  and\ \bibinfo {author} {\bibfnamefont {T.}~\bibnamefont {Baumbach}},\
  }\href@noop {} {\emph {\bibinfo {title} {High-Resolution X-Ray Scattering
  from Thin Films and Multilayers}}},\ Springer Tracts in Modern Physics\
  (\bibinfo  {publisher} {Springer Berlin Heidelberg},\ \bibinfo {year}
  {1999})\BibitemShut {NoStop}%
\bibitem [{\citenamefont {Daillant}\ and\ \citenamefont
  {Gibaud}(2008)}]{Daillant_2008}%
  \BibitemOpen
  \bibfield  {author} {\bibinfo {author} {\bibfnamefont {J.}~\bibnamefont
  {Daillant}}\ and\ \bibinfo {author} {\bibfnamefont {A.}~\bibnamefont
  {Gibaud}},\ }\href {https://books.google.de/books?id=fk5tCQAAQBAJ} {\emph
  {\bibinfo {title} {X-ray and Neutron Reflectivity: Principles and
  Applications}}},\ Lecture Notes in Physics\ (\bibinfo  {publisher} {Springer
  Berlin Heidelberg},\ \bibinfo {year} {2008})\BibitemShut {NoStop}%
\bibitem [{\citenamefont {Durand}\ and\ \citenamefont
  {Morizet}(2006)}]{Durand_2006}%
  \BibitemOpen
  \bibfield  {author} {\bibinfo {author} {\bibfnamefont {O.}~\bibnamefont
  {Durand}}\ and\ \bibinfo {author} {\bibfnamefont {N.}~\bibnamefont
  {Morizet}},\ }\bibfield  {title} {\enquote {\bibinfo {title}
  {Fourier-inversion and wavelet-transform methods applied to x-ray
  reflectometry and hrxrd profiles from complex thin-layered
  heterostructures},}\ }\href {\doibase
  https://doi.org/10.1016/j.apsusc.2006.05.106} {\bibfield  {journal} {\bibinfo
   {journal} {Applied Surface Science}\ }\textbf {\bibinfo {volume} {253}},\
  \bibinfo {pages} {133 -- 137} (\bibinfo {year} {2006})},\ \bibinfo {note}
  {proceedings of the E-MRS 2005 Spring Meeting Symposium P: Current trends in
  optical and X-ray metrology of advanced materials for nanoscale
  devices}\BibitemShut {NoStop}%
\bibitem [{\citenamefont {Sakurai}, \citenamefont {Mizusawa},\ and\
  \citenamefont {Ishii}(2008)}]{Sakurai_2008}%
  \BibitemOpen
  \bibfield  {author} {\bibinfo {author} {\bibfnamefont {K.}~\bibnamefont
  {Sakurai}}, \bibinfo {author} {\bibfnamefont {M.}~\bibnamefont {Mizusawa}}, \
  and\ \bibinfo {author} {\bibfnamefont {M.}~\bibnamefont {Ishii}},\ }\bibfield
   {title} {\enquote {\bibinfo {title} {Significance of frequency analysis in
  x-ray rflectivity: Towards analysis which does not depend too much on
  models},}\ }\href {\doibase 10.14723/tmrsj.33.523} {\bibfield  {journal}
  {\bibinfo  {journal} {Transactions of the Materials Research Society of
  Japan}\ }\textbf {\bibinfo {volume} {33}},\ \bibinfo {pages} {523--528}
  (\bibinfo {year} {2008})}\BibitemShut {NoStop}%
\bibitem [{\citenamefont {Durand}(2004)}]{Durand_2004}%
  \BibitemOpen
  \bibfield  {author} {\bibinfo {author} {\bibfnamefont {O.}~\bibnamefont
  {Durand}},\ }\bibfield  {title} {\enquote {\bibinfo {title} {Characterization
  of multilayered materials for optoelectronic components by high-resolution
  x-ray diffractometry and reflectometry: contribution of numerical
  treatments},}\ }\href {\doibase https://doi.org/10.1016/j.tsf.2003.10.052}
  {\bibfield  {journal} {\bibinfo  {journal} {Thin Solid Films}\ }\textbf
  {\bibinfo {volume} {450}},\ \bibinfo {pages} {51 -- 59} (\bibinfo {year}
  {2004})},\ \bibinfo {note} {proceedings of Symposium M on Optical and X-Ray
  Metrology for Advanced Device Materials Characterization, of the E-MRS 2003
  Spring Conference}\BibitemShut {NoStop}%
\bibitem [{\citenamefont {Donnelle}\ and\ \citenamefont
  {Rust}(2005)}]{Donelle_2005}%
  \BibitemOpen
  \bibfield  {author} {\bibinfo {author} {\bibfnamefont {D.}~\bibnamefont
  {Donnelle}}\ and\ \bibinfo {author} {\bibfnamefont {B.}~\bibnamefont
  {Rust}},\ }\bibfield  {title} {\enquote {\bibinfo {title} {The fast fourier
  transform for experimentalists. part i. concepts},}\ }\href@noop {}
  {\bibfield  {journal} {\bibinfo  {journal} {Computing in Science
  Engineering}\ }\textbf {\bibinfo {volume} {7}},\ \bibinfo {pages} {80--88}
  (\bibinfo {year} {2005})}\BibitemShut {NoStop}%
\bibitem [{\citenamefont {Davor}(1993)}]{Balzar_1993}%
  \BibitemOpen
  \bibfield  {author} {\bibinfo {author} {\bibfnamefont {B.}~\bibnamefont
  {Davor}},\ }\bibfield  {title} {\enquote {\bibinfo {title} {X-ray diffraction
  line broadening: Modeling and applications to high-tc superconductors},}\
  }\href {https://doi.org/10.6028/jres.098.026} {\bibfield  {journal} {\bibinfo
   {journal} {Journal of research of the National Institute of Standards and
  Technology}\ }\textbf {\bibinfo {volume} {98}},\ \bibinfo {pages} {321–353}
  (\bibinfo {year} {1993})}\BibitemShut {NoStop}%
\bibitem [{\citenamefont {Sevenhans}\ \emph {et~al.}(1986)\citenamefont
  {Sevenhans}, \citenamefont {Gijs}, \citenamefont {Bruynseraede},
  \citenamefont {Homma},\ and\ \citenamefont {Schuller}}]{Sevenhans_1986}%
  \BibitemOpen
  \bibfield  {author} {\bibinfo {author} {\bibfnamefont {W.}~\bibnamefont
  {Sevenhans}}, \bibinfo {author} {\bibfnamefont {M.}~\bibnamefont {Gijs}},
  \bibinfo {author} {\bibfnamefont {Y.}~\bibnamefont {Bruynseraede}}, \bibinfo
  {author} {\bibfnamefont {H.}~\bibnamefont {Homma}}, \ and\ \bibinfo {author}
  {\bibfnamefont {I.~K.}\ \bibnamefont {Schuller}},\ }\bibfield  {title}
  {\enquote {\bibinfo {title} {Cumulative disorder and x-ray line broadening in
  multilayers},}\ }\href {\doibase 10.1103/PhysRevB.34.5955} {\bibfield
  {journal} {\bibinfo  {journal} {Phys. Rev. B}\ }\textbf {\bibinfo {volume}
  {34}},\ \bibinfo {pages} {5955--5958} (\bibinfo {year} {1986})}\BibitemShut
  {NoStop}%
\bibitem [{\citenamefont {Delhez}, \citenamefont {de~Keijser},\ and\
  \citenamefont {Mittemeijer}(1978)}]{Delhez_1978}%
  \BibitemOpen
  \bibfield  {author} {\bibinfo {author} {\bibfnamefont {R.}~\bibnamefont
  {Delhez}}, \bibinfo {author} {\bibfnamefont {T.~H.}\ \bibnamefont
  {de~Keijser}}, \ and\ \bibinfo {author} {\bibfnamefont {E.~J.}\ \bibnamefont
  {Mittemeijer}},\ }\bibfield  {title} {\enquote {\bibinfo {title} {The x-ray
  diffraction line broadening due to the diffractometer condition as a function
  of 2$\theta$},}\ }\href {\doibase 10.1088/0022-3735/11/7/015} {\bibfield
  {journal} {\bibinfo  {journal} {Journal of Physics E: Scientific
  Instruments}\ }\textbf {\bibinfo {volume} {11}},\ \bibinfo {pages} {649--652}
  (\bibinfo {year} {1978})}\BibitemShut {NoStop}%
\bibitem [{\citenamefont {Sinha}\ \emph {et~al.}(1988)\citenamefont {Sinha},
  \citenamefont {Sirota}, \citenamefont {Garoff},\ and\ \citenamefont
  {Stanley}}]{Sinha_1988}%
  \BibitemOpen
  \bibfield  {author} {\bibinfo {author} {\bibfnamefont {S.~K.}\ \bibnamefont
  {Sinha}}, \bibinfo {author} {\bibfnamefont {E.~B.}\ \bibnamefont {Sirota}},
  \bibinfo {author} {\bibfnamefont {S.}~\bibnamefont {Garoff}}, \ and\ \bibinfo
  {author} {\bibfnamefont {H.~B.}\ \bibnamefont {Stanley}},\ }\bibfield
  {title} {\enquote {\bibinfo {title} {X-ray and neutron scattering from rough
  surfaces},}\ }\href {\doibase 10.1103/PhysRevB.38.2297} {\bibfield  {journal}
  {\bibinfo  {journal} {Phys. Rev. B}\ }\textbf {\bibinfo {volume} {38}},\
  \bibinfo {pages} {2297--2311} (\bibinfo {year} {1988})}\BibitemShut {NoStop}%
\bibitem [{\citenamefont {Lammel}\ \emph {et~al.}(2020)\citenamefont {Lammel},
  \citenamefont {Geishendorf}, \citenamefont {Choffel}, \citenamefont {Hamann},
  \citenamefont {Johnson}, \citenamefont {Nielsch},\ and\ \citenamefont
  {Thomas}}]{code_DOI}%
  \BibitemOpen
  \bibfield  {author} {\bibinfo {author} {\bibfnamefont {M.}~\bibnamefont
  {Lammel}}, \bibinfo {author} {\bibfnamefont {K.}~\bibnamefont {Geishendorf}},
  \bibinfo {author} {\bibfnamefont {M.~A.}\ \bibnamefont {Choffel}}, \bibinfo
  {author} {\bibfnamefont {D.~M.}\ \bibnamefont {Hamann}}, \bibinfo {author}
  {\bibfnamefont {D.~C.}\ \bibnamefont {Johnson}}, \bibinfo {author}
  {\bibfnamefont {K.}~\bibnamefont {Nielsch}}, \ and\ \bibinfo {author}
  {\bibfnamefont {A.}~\bibnamefont {Thomas}},\ }\bibfield  {title} {\enquote
  {\bibinfo {title} {Fft-multi-gaussian-fitting-routine.ipynb [python 3.x]},}\
  }\href@noop {} {\  (\bibinfo {year} {2020})},\ \Eprint
  {http://arxiv.org/abs/arXiv:2008.04626} {arXiv:2008.04626} \BibitemShut
  {NoStop}%
\bibitem [{\citenamefont {Wang}\ \emph {et~al.}(1998)\citenamefont {Wang},
  \citenamefont {Weiss}, \citenamefont {Shaikhutdinov}, \citenamefont {Ritter},
  \citenamefont {Petersen}, \citenamefont {Wagner}, \citenamefont {Schl\"ogl},\
  and\ \citenamefont {Scheffler}}]{Wang_1998}%
  \BibitemOpen
  \bibfield  {author} {\bibinfo {author} {\bibfnamefont {X.-G.}\ \bibnamefont
  {Wang}}, \bibinfo {author} {\bibfnamefont {W.}~\bibnamefont {Weiss}},
  \bibinfo {author} {\bibfnamefont {S.~K.}\ \bibnamefont {Shaikhutdinov}},
  \bibinfo {author} {\bibfnamefont {M.}~\bibnamefont {Ritter}}, \bibinfo
  {author} {\bibfnamefont {M.}~\bibnamefont {Petersen}}, \bibinfo {author}
  {\bibfnamefont {F.}~\bibnamefont {Wagner}}, \bibinfo {author} {\bibfnamefont
  {R.}~\bibnamefont {Schl\"ogl}}, \ and\ \bibinfo {author} {\bibfnamefont
  {M.}~\bibnamefont {Scheffler}},\ }\bibfield  {title} {\enquote {\bibinfo
  {title} {The hematite ( $\mathit{\ensuremath{\alpha}}$-
  ${\mathrm{fe}}_{2}{O}_{3}$) (0001) surface: Evidence for domains of distinct
  chemistry},}\ }\href {\doibase 10.1103/PhysRevLett.81.1038} {\bibfield
  {journal} {\bibinfo  {journal} {Phys. Rev. Lett.}\ }\textbf {\bibinfo
  {volume} {81}},\ \bibinfo {pages} {1038--1041} (\bibinfo {year}
  {1998})}\BibitemShut {NoStop}%
\bibitem [{\citenamefont {Madelung}(2000)}]{Landolt-Boernstein_hematite}%
  \BibitemOpen
  \bibinfo {editor} {\bibfnamefont {O.}~\bibnamefont {Madelung}},\ ed.,\
  \enquote {\bibinfo {title} {Hematite (alpha-fe2o3): general characterization,
  crystal structure, lattice parameters},}\ in\ \href {\doibase
  10.1007/10681735_543} {\emph {\bibinfo {booktitle} {Landolt-B{\"o}rnstein -
  Group III Condensed Matter}}},\ Vol.\ \bibinfo {volume} {41D}\ (\bibinfo
  {publisher} {Springer-Verlag Berlin Heidelberg},\ \bibinfo {year}
  {2000})\BibitemShut {NoStop}%
\bibitem [{\citenamefont {Hellwege}(1975)}]{Landolt-Boernstein_yttria}%
  \BibitemOpen
  \bibinfo {editor} {\bibfnamefont {K.-H.}\ \bibnamefont {Hellwege}},\ ed.,\
  \enquote {\bibinfo {title} {b172, ii.1.1 simple oxides},}\ in\ \href
  {\doibase 10.1007/10201470_5} {\emph {\bibinfo {booktitle}
  {Landolt-B{\"o}rnstein - Group III Condensed Matter}}},\ Vol.\ \bibinfo
  {volume} {7B1}\ (\bibinfo  {publisher} {Springer-Verlag Berlin Heidelberg},\
  \bibinfo {year} {1975})\BibitemShut {NoStop}%
\end{thebibliography}%

\end{document}